\documentclass[aps,pre,groupedaddress,preprint]{revtex4}
\usepackage{graphicx}
\usepackage[dvips]{epsfig}
\usepackage{dcolumn}
\usepackage{subfigure}%
\usepackage{bm}
\usepackage{amssymb}
\usepackage{amsmath}

\begin{document}
\title{Dynamics of Generalized Hydrodynamics: Hyperbolic and Pseudohyperbolic Burgers Equations}

\author{Carlos Escudero}

\affiliation{ICMAT (CSIC-UAM-UC3M-UCM), Departamento de
Matem\'{a}ticas, Facultad de Ciencias, Universidad Aut\'{o}noma de
Madrid, Ciudad Universitaria de Cantoblanco, 28049 Madrid, Spain}

\begin{abstract}
The equations of continuum hydrodynamics can be derived from the
Boltzmann equation, which describes rarefied gas dynamics at the
kinetic level, by means of the Chapman-Enskog expansion. This
expansion assumes a small Knudsen number, and as a consequence,
the hydrodynamics equations are able to successfully describe
sound propagation when the frequency of a sound wave is much
higher than the collision frequency of the particles. When both
frequencies become comparable, these equations give a poor account
of the experimental measurements. A series of generalized
hydrodynamic equations has been introduced in the literature along
the years in order to improve the continuous description of small
scale properties of fluid flow, as ultrasound propagation. We will
describe herein some of the simplified models that has been
proposed so far.
\end{abstract}

\maketitle

One of the main challenges of nonequilibrium statistical mechanics
is to obtain accurate macroscopic descriptions of microscopic
processes. An important example is the derivation of the
hydrodynamics, Euler and Navier–-Stokes, equations from the
Boltzmann equation~\cite{bardos,raymond}, which describes rarefied
gas dynamics at the kinetic level. This program can be carried out
by means of the Chapman–-Enskog expansion~\cite{chapman,carlo},
assuming a small Knudsen number, which is the ratio among the mean
free path of the gas molecules and the macroscopic characteristic
length. While the Navier-–Stokes equations represent a remarkable
success of the theoretical study of fluid mechanics, it is well
known that the spectral properties of their solution do not agree
with experimental data for short wavelengths~\cite{rao}.
Consequently, the Navier--Stokes equations are able to
successfully describe sound propagation when the frequency of a
sound wave is much higher than the collision frequency of the
particles. When both frequencies become comparable, these
equations give a poor account of the experimental measurements. In
order to solve this problem, one is tempted to derive generalized
hydrodynamic equations continuing the Chapman–-Enskog expansion to
higher orders~\cite{burnett}, to obtain the so called Burnett and
supra--Burnett equations, but they have never achieved any notable
success~\cite{grad}.

Of course, studying and analyzing the Euler, Navier--Stokes,
Burnett and supra--Burnett equations is rather complicated as they
are composed of a large number of terms. This fact justified the
introduction of several simplified models for hydrodynamics. One
of them is the inviscid Burgers equation
\begin{equation}
\partial_t u + u \partial_x u = 0,
\end{equation}
which describes the evolution of the pointwise fluid velocity
$u=u(x,t)$. This equation expresses the free flight of the fluid
particles, and experience non--existence of the solution, in the
form of finite time shock and rarefaction waves~\cite{evans}.
Actual particles do interact, and this interaction can be
introduced phenomenologically into the equation for fluid motion
by means of a viscosity term, to obtain the viscous Burgers
equation
\begin{equation}
\partial_t u + u \partial_x u = \nu \partial_x^2 u,
\end{equation}
where $\nu$ is the fluid viscosity. This equation can be
explicitly solved by means of the Hopf--Cole
transformation~\cite{evans}, and the exact formula reveals that
the solution is regular for all times. However, contrarily to what
happens in the inviscid case, in which perturbations propagate
linearly in time, the viscous case supports infinitely fast
propagation of disturbances. The hyperbolic modification of the
Burgers equation
\begin{equation}
\mu \partial_t^2 u + \partial_t u + u \partial_x u = \nu
\partial_x^2 u,
\end{equation}
where $\mu$ is the fluid inertia, can be introduced
phenomenologically in order to take into account the memory
effects coming from the finite size of the temporal lapse among
gas molecules collisions. One would in principle expect that such
a wave form of the hyperbolic Burgers equation is able to support
finite propagation of disturbances, a fact that can also be proven
rigourously~\cite{escudero}.

Beyond phenomenology, it is possible to find more fundamental
justifications of the wave form of the hyperbolic Burgers
equation. One possibility was suggested by Rosenau~\cite{rosenau},
who found that the telegraphers equation (which is actually the
linearization about $u=0$ of the hyperbolic Burgers equation)
\begin{equation}
\mu \partial_t^2 u + \partial_t u = \nu
\partial_x^2 u,
\end{equation}
reproduces the spectrum of its microscopic counterpart, the
persistent random walk, almost exactly. Using this fact, he
claimed that the Chapman–-Enskog expansion should be substituted
by a different expansion keeping space and time on equal footing.
This procedure would preserve the hyperbolic nature of the
resulting equations, and thus the nice spectral properties of the
solution, at least in the linear regime~\cite{rosenau}. An
expansion of this type was carried out by Khonkin~\cite{khonkin},
who found new equations for the momentum and energy fluxes which,
in contrast to the classical Navier–-Stokes and Fourier laws,
depend on the first time derivative of these fluxes. This
dependence implies in turn the appearance of a term proportional
to the second order time derivative of the velocity, among others,
in the corresponding modified Navier–-Stokes equations, which
become now hyperbolic. However, it was already argued by Rosenau
that hyperbolicity in union with nonlinear hydrodynamical
evolution might result in the non--existence of the solution. A
similar idea was proposed later~\cite{makarenko}, where
hyperbolicity was introduced to take into account memory effects
in the hydrodynamic description of the flow, and to get rid of the
infinite speed of signal propagation. In order to understand the
interplay between hyperbolicity and nonlinear convection, the
hyperbolic Burgers equation was studied in~\cite{makarenko} by
means of linear and numerical analyses. One of the conclusions of
this work is that this equation has blowing up solutions under
certain circumstances. Note also that, apart from its use as a toy
model for generalized hydrodynamics, the hyperbolic Burgers
equation has been employed as mathematical model of traffic
flow~\cite{lighthill,jordan}.

The hyperbolic Burgers equation has been rigourously analyzed a
number of times. In~\cite{rudjak}, the global existence in time of
the solution was proved for small enough initial conditions,
together with the convergence of this solution to the
corresponding one of the viscous Burgers equation in the limit
$\mu \to 0$. The non-existence of a global in time solution of a
certain class of nonlinear wave equations was proven
in~\cite{sideris}; the hyperbolic Burgers equation can be shown to
belong to this class. The temporal asymptotic behavior of the
solution to the hyperbolic Burgers equation was proved
in~\cite{zuazua} to be the same as the one of the viscous Burgers
equation, in those cases in which this solution is global in time.
The shock wave dynamics of initial discontinuous profiles was
studied in~\cite{jordan,jordan2} by means of singular surface
theory and numerical approximations. In~\cite{escudero} we proved
the finite time blow-up of the solution to the hyperbolic Burgers
equation provided its initial conditions where large enough. We
showed that for regular, large enough and compactly supported
initial conditions the solution $u$ to the hyperbolic Burgers
equation obeys
\begin{equation}
\lim_{t \to t_b} \left| \left| u \right|
\right|_{L^\infty(\mathbb{R})} = \infty,
\end{equation}
for some finite $t_b$. We have proven additionally some other
properties of the solution to the hyperbolic Burgers equation. It
cannot develop any other form of non-existence of the solution,
apart from blow-up, provided the initial conditions are regular
enough. This prohibits in particular the finite time formation of
shock or acceleration waves out of regular initial data. In more
general terms we can say that we have proven that the solution is
regular and compactly supported as long as it is bounded.
See~\cite{escudero} for details of the proofs.

One sees that after introducing inertia in the viscous Burgers
equation to get the hyperbolic Burgers equation one recovers both
the desirable finite speed of propagation of disturbances and the
undesirable finite time non-existence of the solution. To prevent
the singularities a higher order viscosity term was introduced
in~\cite{makarenko}, and so the pseudohyperbolic Burgers equation
\begin{equation}
\mu \partial_t^2 u + \partial_t u + u \partial_x u = \nu
\partial_x^2 u + \lambda \partial_x^2 \partial_t u,
\end{equation}
was found. Note that herein we have assumed a constant
hyperviscosity $\lambda > 0$, contrary to the $x-$dependent one
in~\cite{makarenko}, which sign was not defined either (in this
last case the equation becomes linearly ill-posed). Contrary to
the conjecture raised in~\cite{makarenko}, our preliminary
numerical analyses~\cite{escuderojordan} have shown that the
solution to the pseudohyperbolic Burgers equations, which
propagates perturbations infinitely fast as one could na\"{\i}fly
expect, also blows up in finite time provided the initial
conditions are large enough. In this case, however, blow-up
requires much larger initial conditions than in the case of the
hyperbolic Burgers equation. In this sense we can say that
hyperviscosity implies a partial regularization of the solution,
but it still unable to unconditionally prevent blow-up.

\section*{Acknowledgments}

The author is grateful to Pedro Jordan for his interest in our
work and for sharing with us his knowledge on this topic. Our
joint numerical investigation of the pseudohyperbolic Burgers
equation was performed during the conference ICTCA 2009 in
Dresden. This work has been partially supported by the MICINN
(Spain) through Project No. MTM2008-03754.

\end{document}